\documentclass[aps,prb,showpacs, groupedaddress,twocolumn, superscriptaddress]{revtex4}
\usepackage{amsmath,graphicx,latexsym,times,color}
\usepackage{setspace}
\usepackage{hyperref}
\usepackage{array}
\usepackage{titlesec}

\begin{document}

\title{Perfect spin filtering by symmetry in molecular junctions}

\author{Dongzhe Li}
\affiliation{Service de Physique de l'Etat Condens\'e (SPEC), CEA, CNRS, Universit\'e Paris-Saclay, CEA Saclay 91191 Gif-sur-Yvette Cedex, France}
\affiliation{Department of Physics, University of Konstanz, 78457 Konstanz, Germany}

\author{Yannick J. Dappe}
\affiliation{Service de Physique de l'Etat Condens\'e (SPEC), CEA, CNRS, Universit\'e Paris-Saclay, CEA Saclay 91191 Gif-sur-Yvette Cedex, France}

\author{Alexander Smogunov}
\email{alexander.smogunov@cea.fr}
\affiliation{Service de Physique de l'Etat Condens\'e (SPEC), CEA, CNRS, Universit\'e Paris-Saclay, CEA Saclay 91191 Gif-sur-Yvette Cedex, France}

\date{\today}

\begin{abstract}
Obtaining highly spin-polarized currents in molecular junctions is crucial and desirable for nanoscale spintronics devices. Motivated by our recent symmetry-based theoretical argument for complete blocking of one spin conductance channel in atomic-scale junctions [A. Smogunov and Y. J. Dappe, Nano Lett. 15, 3552 (2015)], we explore the generality of the proposed mechanism and the degree of achieved spin-polarized current for various ferromagnetic electrodes (Ni, Co, Fe) and two different molecules, quaterthiophene and p-quaterphenyl. A simple analysis of the spin-resolved local density of states of a free electrode allowed us to identify the Fe(110) as the most optimal electrode, providing perfect spin filtering and high conductance at the same time. These results are confirmed by $ab$ $initio$ quantum transport calculations and are similar to those reported previously for model junctions. It is found, moreover, that the distortion of the p-quaterphenyl molecule plays an important role, reducing significantly the overall conductance.
\end{abstract}


\newcommand{\Ea}{\ensuremath{E_a}}
\newcommand{\Eb}{\ensuremath{E_b}}
\newcommand{\Ec}{\ensuremath{E_c}}
\newcommand{\Ed}{\ensuremath{E_d}}
\newcommand{\Ee}{\ensuremath{E_e}}

\newcommand{\eV}{\ensuremath{\,eV}}

\maketitle

Molecular spintronics is a very promising and emerging field, whose aim is to manipulate the spin degree of freedom in molecular based devices \cite{Rocha_2005, Sanvito-2010}.  Such devices
should possess a large spin-relaxation length which is important for using spin degree of freedom in transport properties. In particular, due to low spin-orbit coupling in organic molecules, the electron spin-relaxation length will be rather large, making
such organic-based devices very promising for spintronics applications.
The property of high interest in spintronics is the spin-filtering or the spin polarization of (zero bias) electric current, $\text{P}=(G_{\uparrow}-G_{\downarrow})/(G_{\uparrow}+G_{\downarrow})$, where $G_{\uparrow}$ and $G_{\downarrow}$ are the conductances of majority and minority spin channels, respectively. 
Another important property is the magneto-resistance (MR) - the strong change in electrical conductance $G$ between parallel and antiparallel magnetic alignments of two ferromagnetic electrodes - $\text{MR}=(G_{\text P}-G_{\text {AP}})/G_{\text {AP}}$. In this context, achieving as large as possible P (ideally, 100\%) and MR (ideally, infinite) represents a crucial issue.

Unexpected large MR of up to 300\% and large spin-dependent transport lengths have been reported in spin valves using Alq$_3$ \cite{Barraud-2010, Sun-2010}. The MR of about 60\%, 67\% and
80\% have been reported for Co$|$C$_{60}$$|$Co \cite{Xiangmin-2015}, Fe$|$C$_{70}$$|$Fe \cite{C70-2014} and Ni$|$C$_{60}$$|$Ni \cite{Kenji-2013} magnetic junctions, respectively. At the
single molecular scale, due to strong hybridization between molecular orbitals and ferromagnetic electrode states, it has been found that the giant MR (GMR) can reach rather large values,
ranging from 50\% to 80\% \cite{Schmaus-2011, Alexei-2012, Tao-2013}, in the contact regime. In the tunneling regime, a tunneling MR (TMR) value of up to 100\% has been found in the case of
single C$_{60}$ molecule deposited on the Chromium surface \cite{Alex-2012}. Moreover, a selective and efficient spin injection at the ferromagnetic-organic interface has also been reported
by locally controlling the inversion of the spin polarization close to the Fermi level \cite{Atodiresei-2010}. This effect, however, was found to depend strongly on the details of
nanocontact configuration as well as on the molecule-electrode coupling.

Quite generally, the conductance in ferromagnetic nanocontacts is dominated by weakly polarized $s$ orbitals resulting in a partial spin-polarized current 
\cite{Jacob-2005, Smogunov-2006, Cuevas-2008}. 
Therefore, blocking transport via $s$ orbitals and promoting transport via $d$ (or $p$) orbitals seems to be a good strategy to obtain high spin-polarizations. Very recently, a 100\% spin-polarized currents have been reported experimentally in the nickel oxide atomic junctions formed between two nickel electrodes \cite{Vardimon2015} by the break-junction setup. The current work is motivated by our recent study of spin-polarized electron transport through a special class of $\pi$-conjugated molecules bridging two Ni electrodes \cite{Alex_2015}. 
Due to symmetry mismatch between molecular orbitals and conductance channels of the Ni electrodes, the electron transport was carried by only $d$-orbitals while the $s$-conductance channels were fully blocked at the electrode-molecule junction. 
As a consequence, a perfect 100\% spin-polarization ($G_{\downarrow} \approx 0.65 G_0, G_{\uparrow} = 0$) and an infinite MR have been found for ideal Ni electrodes (represented by 
semi-infinite Ni chains), while rather moderate values, $G_{\downarrow} \approx 0.19 G_0, G_{\uparrow} \approx 0.02 G_0$, were found for more realistic Ni(111) electrodes.

The purpose of this Rapid Communication is two-fold:
i) to identify molecular bridges (the electrodes and the molecule) employing in full our symmetry argument 
for perfect spin-filtering; 
ii) to demonstrate the generality of our symmetry-mismatch mechanism. It will be shown that the Fe(110) electrode is the optimal one, providing highest spin-polarization of incoming conductance channels while polythiophene molecules are probably the best ones due to their highest occupied molecular orbital (HOMO) placed very closely to the Fermi level. 

\begin{figure*}[!htbp]
\centering
\includegraphics[scale=0.55]{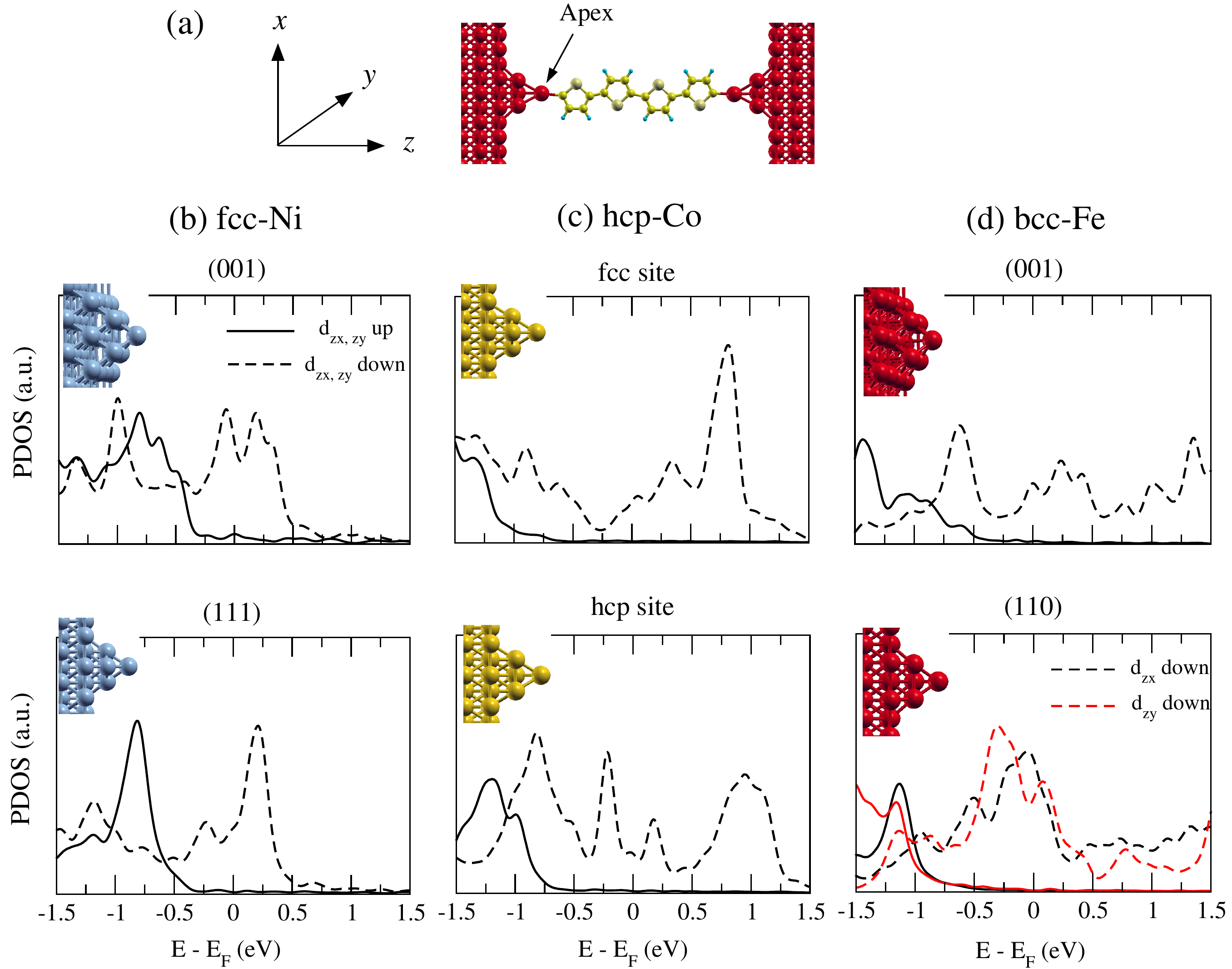}
\caption{\label{PDOS-Apex}
(Color online) (a) The model geometry of ferromagnetic$|$molecule$|$ferromagnetic nanojunction used in the paper. 
Spin-polarized projected density of states (PDOS) on $d_{zx, zy}$ orbitals of one of (equivalent) apex atoms of the electrodes for 
(b) fcc-Ni, of (001) (top) and (111) (bottom) orientations;  
(c) hcp-Co, of fcc (top) and hcp (bottom) positions of the pyramid-like tip;  
(d) bcc-Fe, of (001) (top) and (110) (bottom) orientations. 
The insets show the atomic structures of corresponding electrodes which are terminated 
with only one apex atom. The pyramids contain either 4 [Ni(111), hcp-Co, and Fe(110)] or 5 [Ni(001) and Fe(001)] atoms. 
The $d_{zx}$ and $d_{zy}$ orbitals are degenerate for all presented electrodes except for Fe(110) due to symmetry reason. 
Spin up (down) results are plotted by solid (dashed) lines.
}
\end{figure*}

{\it DFT-based spin-dependent quantum transport.} The spin-polarized $\textit {ab initio}$ calculations were carried out using the plane wave electronic structure package QUANTUM ESPRESSO \cite{Giannozzi2009} in the framework of the density functional theory (DFT). We used the local density approximation (LDA) with Perdew-Zunger parametrization \cite{LDA_PZ} of the exchange-correlation functional with ultra-soft pseudopotentials (USPP) to describe electron-ions interaction. The molecular junctions were simulated by a supercell as the one shown in Fig.\ref{PDOS-Apex}(a). Details concerning the calculations can be found in Ref. \cite{comment}. For the calculations of ballistic transport across the junction, the unit cell of Fig.\ref{PDOS-Apex}(a) was considered as a scattering region joint perfectly 
on both sides to semi-infinite electrodes. Electron transmission was then evaluated using the scattering states approach as implemented in the 
PWCOND code \cite{Alex_PWCOND}. 
Ballistic conductance (at the infinitesimal voltage) is given by Landauer-B\"uttiker formula, $G = G_{0}[T_{\uparrow}(E_{\text{F}})+T_{\downarrow}(E_{\text{F}})]$, where $G_0 = e^2/h$ is
the conductance quantum per spin. A $(8 \times 8)$ $k$-points mesh in the $xy$ plane was found to be enough to obtain well converged transmission functions. 
Notice that the use of the LDA for exchange-correlation potential in a DFT scheme can overestimate the transmission due to the underestimation of band gap \cite{Louie_2007, Thygesen_2011}. However, such effects may affect quantitatively the electron transport properties in our molecular junction, but the main physical trends will remain unaffected due to our robust symmetry argument. Moreover, as we will see later, the conductance is provided by the HOMO molecular orbital whose position with respect to the Fermi level we believe is described quite well by our DFT calculations. 

The key idea of symmetry argument is to block the electrode $s$-channels (both of the majority and minority spins) at the Fermi energy by using $\pi$-conjugated molecules having only $\pi$-states available around the $E_F$. The remaining four transport $d$-channels are all of the minority spin (the majority spin $d$-states are all occupied and lie well below the $E_F$) which would result in a fully spin-polarized current across such molecular junctions. In this context, the spin-polarized density of states at the electrode apex atom plays a crucial role, providing the information on the amount and spin-polarization of states available for the transport. We start therefore by analysing the projected density of states (PDOS) at the electrode apex atom for the free electrodes of different ferromagnetic metals in order to select the best candidate as a spin injection material.
 
{\it Electronic structure of ferromagnetic electrodes.} The ferromagnetic electrodes are modeled by five atomic layers containing 16 atoms in each atomic plane and terminated by a pyramid-like tip. Note that the pyramids contain 4 or 5 atoms for cubic or hexagonal electrodes, respectively  (see insets in the Fig. \ref{PDOS-Apex}). The two bottom layers were fixed while the other three layers and the pyramid were relaxed until the atomic forces are less than 1 meV/\AA.

We emphasize that it is enough to look at only $d_{zx, zy}$ states of the apex atom (Fig. \ref{PDOS-Apex}) since the other apex $d$-states have no overlap with out-of-plane molecular $\pi$-states around the Fermi energy. 
Note that the molecule is assumed to be aligned with the transport direction $Z$ but otherwise can rotate around it. 
In the case if it lies exactly in the $YZ$ plane, for example, the molecular states will hybridize with only $d_{zx}$-states, which are of appropriate symmetry (odd with respect to the molecule $YZ$ plane). We notice also that, by symmetry, the $d_{zx}$ and $d_{zy}$ orbitals are degenerated for all the presented electrodes except for the Fe(110) electrode. 

As expected, a quite general feature has been found for all the ferromagnetic electrodes -- the apex DOS around the Fermi energy is dominated by partially filled minority $d$-states. 
For Ni(001) electrodes (Fig. \ref{PDOS-Apex}(b), upper panel), the spin-down PDOS peak lies in the vicinity of the Fermi energy while the spin-up PDOS is very small but not negligible and starts growing significantly at only $E - E_{\text{F}} < -0.45$ eV. 
In the case of Ni(111) (Fig. \ref{PDOS-Apex}(b), down panel), a relatively smaller [compared to the Ni(001)] spin-down PDOS is found at the Fermi energy
with the maximum around $0.25$ eV above the Fermi energy. 

For hcp-Co electrodes as shown in Fig. \ref{PDOS-Apex}(c), two different adsorption sites, namely fcc and hcp, for the pyramid were considered. 
We found that the hcp site is slightly more favorable with respect to the fcc one, with an energy gain of about 17.84 meV. 
Compared to the fcc-Ni electrode, the apex PDOS for spin down polarization has relatively smaller amplitude at the Fermi energy 
while the spin up states are almost absent. 

For the case of bcc-Fe electrode we have considered two crystallographic orientations, (001) and (110), shown in Fig. \ref{PDOS-Apex}(d) on upper and lower panels,
respectively. The apex atom PDOS of the Fe(001) electrode exhibits a similar spin polarisation as for Ni(111) even though the peaks are rather different. 
On the other hand, almost perfect behavior of the PDOS has been found for the Fe(110) orientation. Namely, rather smooth and large PDOS for spin down states
in a quite large energy window [$-$0.75 eV, $+$0.5 eV] with a maximum around the Fermi energy and almost zero PDOS around the Fermi energy for spin up states. In addition, the non-negligible PDOS for spin up states start about $-0.8$ eV with respect to the Fermi energy.

We conclude, therefore, that among all the electrodes considered, the Fe(110) looks as the optimal one for providing highly spin-polarized incoming current. 
It is expected also to improve considerably over the Ni(111) electrode considered in our previous work.\cite{Alex_2015}

\begin{figure}[!b]
\centering
\includegraphics[scale=0.45]{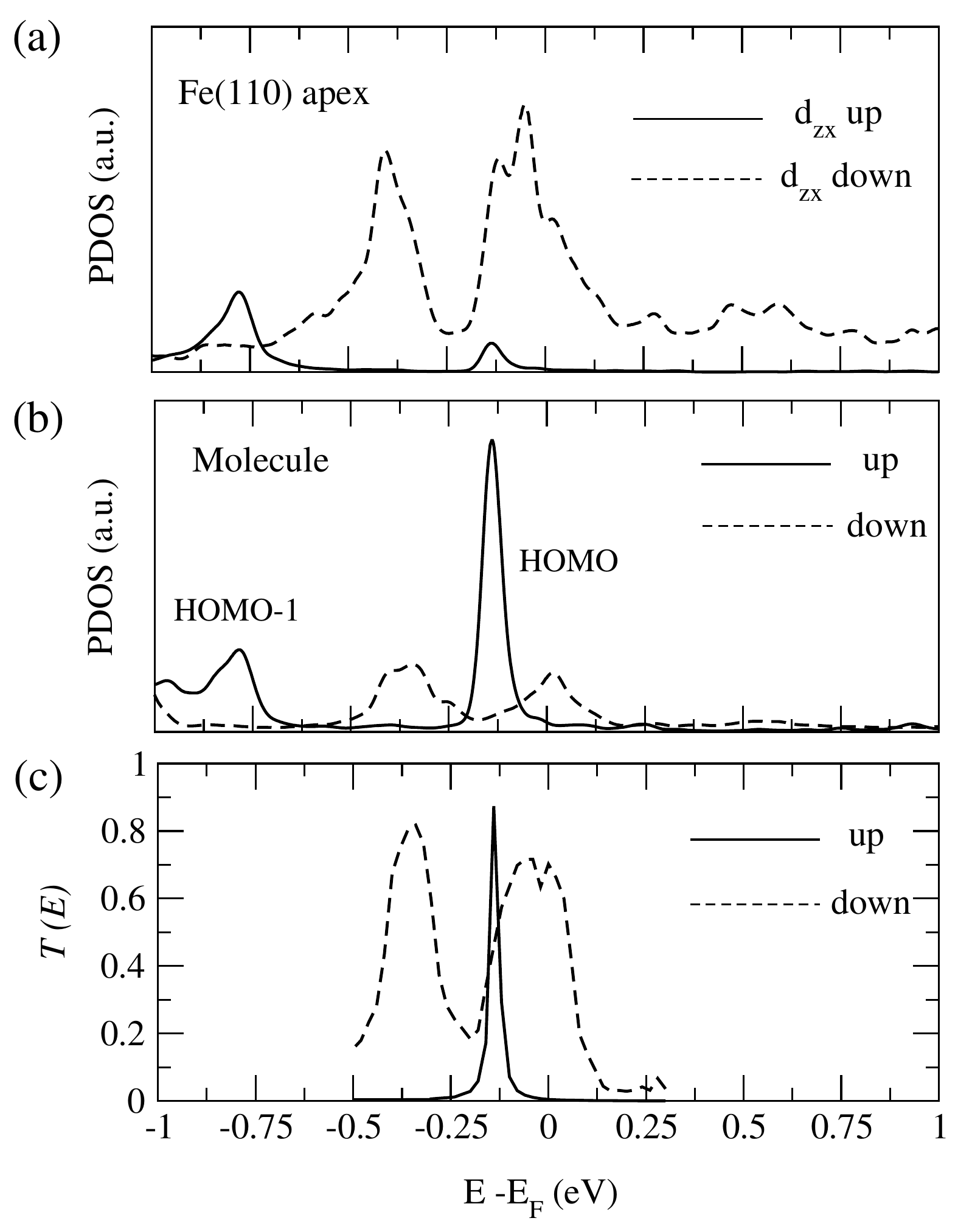}
\caption{\label{Transmission-Fe110}
Electronic and transport properties of Fe(110)$|$quaterthiophene$|$Fe(110) nanojunction: 
(a) spin-dependent PDOS on $d_{zx}$ orbitals of the Fe apex atom, 
(b) spin-resolved density of states on the quaterthiophene molecule, 
(c) spin-resolved transmission function $T(E)$ as a function of the electron energy 
$E$ in the parallel magnetic configuration. 
Spin up (down) results are plotted by solid (dashed) lines.}
\end{figure}

\begin{figure*}[!htbp]
\centering
\includegraphics[scale=0.55]{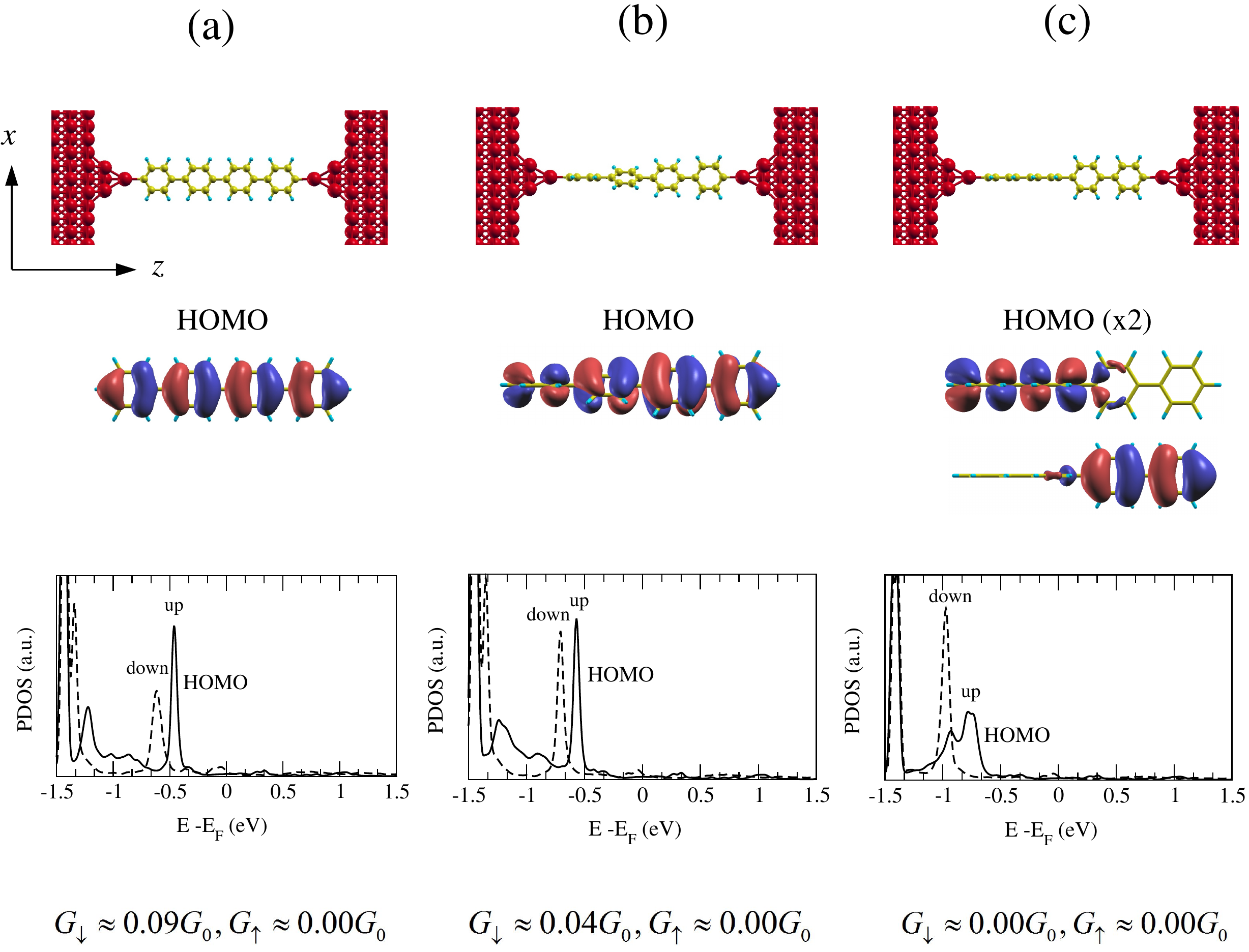}
\caption{\label{Distortion}
(Color online) Electronic and transport properties of p-tetraphenyl molecule suspended between two Fe(110) electrodes.
Three molecular conformations were considered: flat molecule (a), distorted molecule with $30^{\circ}$ rotation between adjacent phenyl cycles (b), 
and distorted molecule with 90$^{\circ}$ rotation (c).
Top panels present model geometries of molecular junctions. 
Middle panels show the charge-density isosurface plots for the HOMO of the free molecule. 
Note that the isosurfaces of positive and negative isovalues are shown in red and blue, respectively. 
Down panels report the spin-resolved PDOS on the molecule, the spin up (down) curves presented by solid (dashed) lines. 
At the bottom, the conductances for both spin channels are indicated.}
\end{figure*}

{\it Fe(110)$|$quaterthiophene$|$Fe(110).} After having selected the best electrode, which in our case turns out to be the Fe(110), we will now confirm our findings by calculating explicitely the spin-polarized transmission functions.
We first consider the quaterthiophene molecule which contains four cycles of thiophene (standing perpendicular to the surface), 
suspended between two Fe(110) electrodes, as shown in Fig. \ref{PDOS-Apex}(a) as a geometric model. 
Experimentally, two well-established approaches, namely the scanning tunneling microscope (STM) tip manipulation and the mechanically controllable break junction (MCBJ) technique, are used in the fabrication of such molecular junctions. For instance, recent experiments showed that such molecular junction could be well established by using the STM tip to pick up one end of 
an individual polythiophene on the surface and subsequently lifting it up \cite{Reecht_2015, Lafferentz_2009}. In addition, the stable and highly conductive molecular junctions were successfully formed by MCBJ technique \cite{Kiguchi2008, Yelin2016}.
By symmetry, the $d_{zx}$-orbital of the Fe apex has non-zero overlap  with the out-of-plane $\pi$-states of the molecule (if the $YZ$ plane is choosen to correspond 
to the molecule plane). The PDOS of the Fe apex, as shown in Fig. \ref{Transmission-Fe110}(a), presents a very similar feature compared to the case of the free electrode 
(see Fig. \ref{PDOS-Apex}(d), down panel). However, a slight splitting of the majority spin states near the Fermi level, as well as a small peak at about $-0.13$ eV for the majority 
spin states arises due to the hybridization with the molecule. Clearly, the PDOS of the Fe apex atom is still dominated by the minority spin states near the Fermi level.

The transport properties of molecular junctions are determined by molecular orbitals located near the Fermi energy. 
In Fig. \ref{Transmission-Fe110}(b) we present the DOS projected on the molecule, 
where one can clearly see that the HOMO is very close to the Fermi energy, meaning that the HOMO is the main transport channel. 

The electronic levels of insulating and semiconducting nanostructure are not described accurately since the DFT is limited to ground states. However, the HOMO described by DFT is equal to the exact ionization potential if we know the exact exchange-correlation potential \cite{Sham1966, Perdew1982}. To address properly the quantum transport in molecular junctions ($i.e.$ the molecule is weakly coupled to the electrodes) one requires here a dynamical treatment of the electron-electron interactions which is beyond the standard DFT mean-field method such as $GW$ approximation. M. Strange {\it et al } \cite{Thygesen_2011} has shown by fully self-consistent $GW$ scheme quantum transport in molecular junctions that the $GW$ is better than DFT for energy alignment of the molecule, however, they found that the difference between $GW$ quasiparticle energies and Kohn-Sham DFT eigenvalues significantly larger for lowest unoccupied molecular orbital (LUMO) than the HOMO. Therefore, the DFT-based quantum transport presented in this work may quantitatively different compared to beyond DFT mean-filed technique, but the main physical trends will remain unaffected.

The spin-up resonance located at about $-0.13$ eV is very sharp due to the lack of appropriate symmetry states in the electrode for the majority spin near the Fermi energy. 
On the contrary, the HOMO state for spin-down channel has significantly broader structure (going from $-0.5$ to $0.13$ eV roughly) 
which reflects the increased hybridization with electrode states. A similar effect (but less pronounced) 
has been also reported for the same molecular junction bridging two Ni(111) electrodes \cite{Alex_2015}. 

As a consequence, the remarquable difference of the conductance for the two spin channels is observed as shown in Fig. \ref{Transmission-Fe110}(c). 
For spin-down electrons, the significant and broad transmission peak is observed. It is located between $-0.1$ eV and $0.1$ eV, 
with a maximum close to the Fermi energy. At the same time, the transmission coefficient in 
the spin-up channel is nearly zero at the Fermi level. 
For that reason, the transport is fully due to spin-down electrons -- 
we find $G_{\downarrow} \approx 0.70 G_0$ and $G_{\uparrow} \approx 0.004 G_0$, for spin down and up polarizations, respectively. 

{\it Fe(110)$|$p-tetraphenyl$|$Fe(110).} In order to understand the effect of a molecule on the spin filtering efficiency, we now investigate another molecular junction made of p-tetraphenyl connecting two Fe(110) electrodes, 
as shown in Fig. \ref{Distortion}. This molecule was chosen because it has much smaller band gap of 0.4 eV in the infinite chain configuration compare to the poly-thiophene chain of 1.4 eV. Therefore, we thought that the corresponding molecular junction will show higher spin down conductance.   

We start by discussing the flat p-tetraphenyl junction as presented in Fig. \ref{Distortion}(a). 
From the charge-density isosurface plots of the HOMO level, we can clearly see the out-of-plane character of the HOMO orbital. 
Compared to the quaterthiophene junction discussed in the previous section, the HOMO was found, however, to be  shifted by about $-0.48$ eV with respect to the Fermi energy, 
which results in the lower spin down conductance, $G_{\downarrow} \approx 0.09 G_0$. The conductance in the spin up channel is again almost zero, indicating a 100\% spin-filtering.

We next investigate the distortion effect of the molecule on its transport property. 
Among three configurations considered (see Fig. \ref{Distortion}), 
we have found that the one with phenyl rings rotated consequenly by 30$^{\circ}$ around the $Z$ axis [Fig. \ref{Distortion}(b)]  
is the most favorable, explicitely, $0.12$ and $0.27$ eV lower in energy than 
the flat molecule [Fig. \ref{Distortion}(a)] and the one with two parts rotated by 90$^{\circ}$ [Fig. \ref{Distortion} (c)], respectively.
Interestingly, the HOMO level moves away from the Fermi energy when the degree of the distortion is increased, {\it i.e.} the HOMO moves to about $-0.52$ and $-0.75$ eV 
for the distortions of 30$^{\circ}$ and 90$^{\circ}$, respectively. 
As a result, a much smaller conductance of $G_{\downarrow} \approx 0.04 G_0$ is found for 30$^{\circ}$ distortion compared to the flat molecule. 
Interestingly, for the 90$^{\circ}$ distortion, we have found almost zero conductance in both spin channels. 
This is attributed to two reasons: i) the 2-fold degenerate HOMOs are strongly localized on either right or left sides [Fig. \ref{Distortion}(c)] 
and are thus completely decoupled from the corresponding electrode, which can be seen as the breaking of the conjugation; 
ii) the HOMO is positioned very far from the Fermi energy. 
In this case, the molecular junction can be seen as two uncoupled molecules sandwiched between the electrodes,
which does not allow the electric current to pass.

{\it Summary.} We present {\it ab initio} quantum transport calculations of spin-polarized electron transport through a special class of $\pi$-conjugated molecules bridging two ferromagnetic 
electrodes. By analysing systematically the PDOS on the apex atom of free ferromagnetic electrodes for different materials as well as for different crystallographic orientations, we
selected the Fe(110) as an optimal electrode for efficient spin injection. Here, a perfect DOS of appropriate symmetry is obtained in spin down channel -- large and smooth around the Fermi
energy --  while spin up contribution is negligibly small. A perfect spin filtering has been found for both quaterthiophene and p-tetraphenyl molecular junctions bridging Fe(110) electrodes
which confirms the generality of our recently proposed symmetry mechanism to obtain fully-polarized currents in single molecule nanojunctions\cite{Alex_2015}. 
Interestingly, a rather high spin down conductance, $G_{\downarrow} \approx 0.70 G_0$, is found for a quaterthiophene junction 
while much smaller value, $G_{\downarrow} \approx 0.09 G_0$, for a p-tetraphenyl one.
Therefore, the quaterthiophene (and, more generally, poly-thiophene molecules) seems to be the best candidate for spin-filtering property, 
which is directly related  to a very close position of its HOMO to the Fermi energy, $E=-0.13$ eV. 
In the case of p-tetraphenyl, the conductance was found to decrease when the phenyl rings are rotated ones with respect to the others 
and gets completely cut in the orthogonal configuration. 
Finally, it should be emphasized that no anti-parallel magnetic alignment of two electrodes were studied explicitely in the paper, however,  
we can argue that the perfectly spin-polarized conductances reported here should directly result also  
in ideally infinite ratios of MR as has been verified in our previous study\cite{Alex_2015}.  
We believe that our results will be important for future electronics and digital information 
technologies based on hybrid metal/organics components.

This work was performed using computation resources from GENCI-[TGCC] project 
(Grant No. 2015097416 and 2016097416).
\\

\bibliographystyle{apsrev}
\bibliography{manuscript}

\begin{thebibliography}{29}
\bibitem[{com()}]{comment}
\bibinfo{note}{The plane wave basis was employed with energy cutoff of 30 and
  300 Ry for the wavefunctions and the charge density, respectively. A $(4
  \times 4)$ in-plane periodicity was used in the $xy$ plane, perpendicular to
  the molecular junctions, in order to minimize unphysical molecule-molecule
  interactions. Four and three metallic layers were employed to model left and
  right electrodes, respectively, and the periodicity in all the three
  directions was assumed in electronic structure calculations. Ionic
  relaxations were carried out using a $(3 \times 3 \times 1)$ $k$-points mesh
  while a more dense $(6 \times 6 \times 1)$ mesh was used for the density of
  states analysis.}
\end{thebibliography}

\end{document}